\documentclass[aps,twocolumn]{revtex4}%
\usepackage{amsfonts}
\usepackage{amsmath}
\usepackage{amssymb}
\usepackage{graphicx}%
\begin{document}

\title[Entanglement Test]{Is state tomography an unambiguous test of quantum
entanglement?}
\author{Niels Gr{\o }nbech-Jensen$^{1}$, Jeffrey E. Marchese$^{1}$, Matteo
Cirillo$^{2}$, and James A. Blackburn$^{3}$ }
\affiliation{$^{1)}$Department of Applied Science, University of California, Davis,
California 95616}
\affiliation{$^{2)}$Dipartimento di Fisica and MINAS-Lab, Universit\`{a} di Roma "Tor
Vergata", I-00173 Roma, Italy}
\affiliation{$^{3)}$Department of Physics \& Computer Science, Wilfrid Laurier
University, Waterloo, Ontario N2L 3C5, Canada}

\begin{abstract}
We provide an alternative interpretation of recently published experimental
results that were represented as demonstrating entanglement between two
macroscopic quantum Josephson oscillators. We model the experimental system
using the well-established classical equivalent circuit of a resistively and
capacitively shunted junction. Simulation results are used to
generate the corresponding density matrix, which is strikingly
similar to the previously published matrix that has been declared to be an
unambiguous demonstration of quantum entanglement. Since our data are
generated by a classical model, we therefore submit that state tomography
cannot be used to determine absolutely whether or not quantum entanglement
has taken place. Analytical arguments are given for why the classical
analysis provides an adequate explanation of the experimental results.
\end{abstract}

\maketitle


The possibility that the supposed quantum behavior of a physical system
might be interpreted within a classical framework has been of interest for
decades. For example, regarding Peierls transitions in quasi one-dimensional
metals, it was suggested that depinning occurring due to quantum tunnelling
could explain experimental results [1], but it was also found that the same
results could likewise be explained by modelling the metal classically with
internal degrees of freedom [2]. Similar observations of dual
interpretations were made in regards to the Pancharatnam-Berry's phase in
optical systems, where an extra geometric phase [3] arises from adiabatic
cyclic evolution of a physical system. In this case the quantum
interpretation of the phase [4] was re-interpreted by several authors [5,6].
With respect to superconducting circuits that include Josephson junctions 
there are two conceptual approaches to interpreting experimental outcomes:
The \textit{classical} Resistively and Capacitively Shunted Junction (RCSJ)
model [7,8], and the quantized model put forward by A.
J. Leggett and co-workers [9]. According to the
latter, under appropriate conditions of temperature and biasing, the phase
difference across a Josephson junction would behave as a macroscopic
quantum variable [10,11,12].
In contrast to the classical RCSJ model, the quantized model thereby
assumes intrinsic discrete energy states in the Josephson potential well, and
escape from the well is possible by quantum tunneling of the phase variable
through the potential barrier.

The first observation of Josephson quantum behavior for a system operated in
washboard potential wells [7,8] was reported in 1985 [11]. Thus began
two decades of investigations of Josephson systems addressing the intriguing
possibility of macroscopic quantum behavior as it would appear with respect
to microwave-induced resonant tunneling, Rabi-oscillations, Ramsey-fringes,
and spin-echo [12]. The essential concept in explaining these experimental
observations was the spacing between the discrete intrinsic energy levels in
the shallow well in combination with the energy carried into the system by
applied microwaves.
The possibility of a dual interpretation of these observations has been
pursued [13] by a systematic reconsideration
of the experiments from the perspective of the RCSJ model. We have found that
the classical description gives a good agreement with reported
experiments on resonant escape, Rabi oscillations, Ramsey fringes, and spin
echo in Josephson systems, all to an impressive degree of accuracy, given
the simplicity of the RCSJ circuit model.
Recently [14] quantum mechanical entanglement was reported
for a system of two weakly coupled superconducting qubits. The significance
of this report is that entanglement is exclusively a quantum
phenomenon. It was claimed that ``A full and unambiguous test of entanglement
comes from state tomography.'' It is that assertion we address in this paper.
More particularly, we ask the question: From the perspective of the
classical model, what would tomography say?


\begin{figure}
[pt]
\begin{center}
\includegraphics[
trim=0.0in 0.0in 0.0in 0.0in,
width=3.20in
]%
{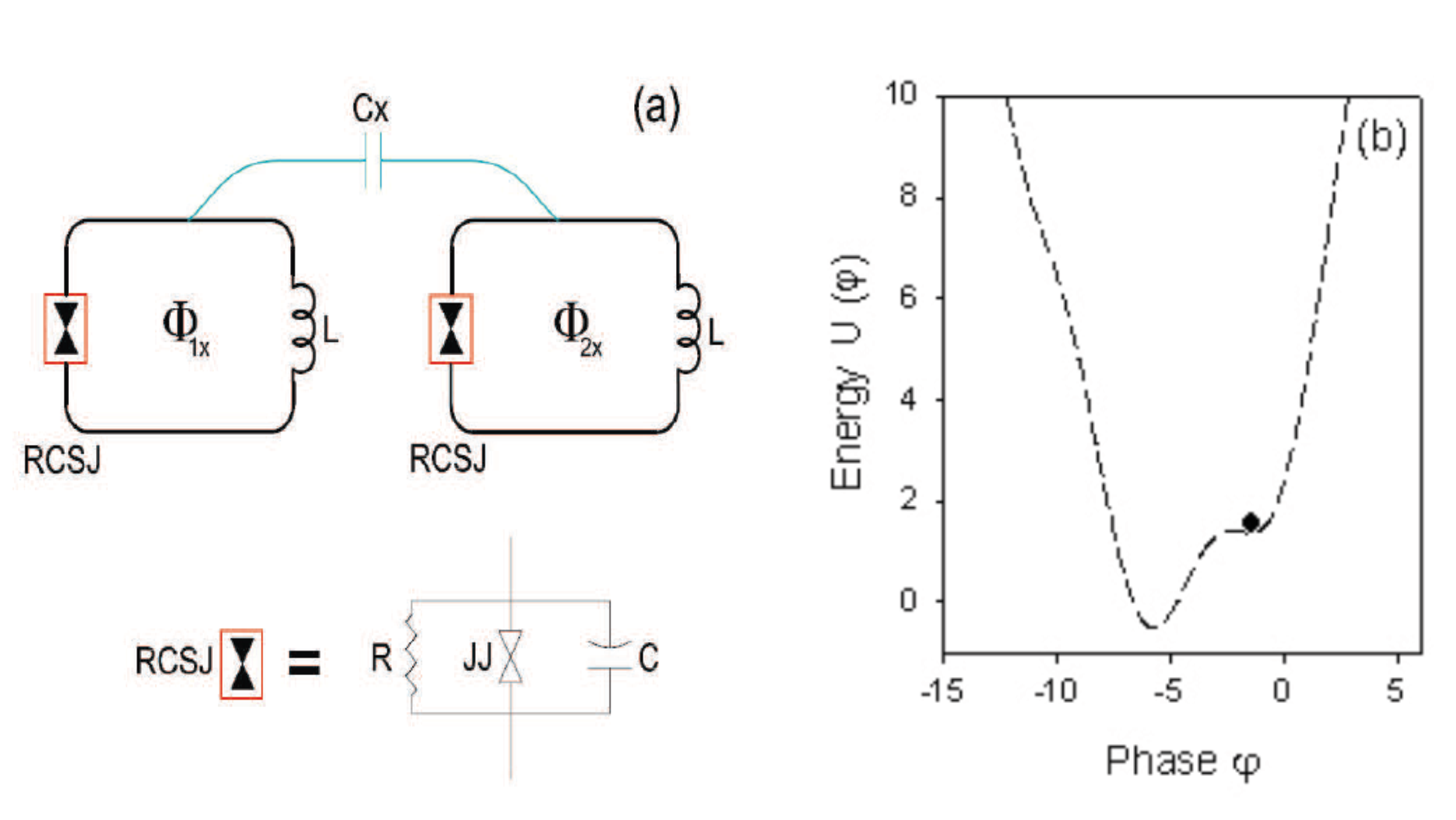}%
\caption{Equivalent circuit of capacitively coupled qubits. Each
superconducting loop of inductance $L$ contains a Josephson junction modeled
by the RCSJ subcircuit. (b) Potential energy (in units of $I_{C}\Phi _{0}/2\pi)$
of a superconducting loop interrupted by a Josephson junction as a
function of the junction phase; the assumed normalized flux bias is 0.6941.
The ``virtual particle'' representing the system is shown sitting at the
bottom of its shallow well.}%
\label{Fig.1}%
\end{center}
\end{figure}

The experimental configuration of two coupled qubits is modeled with
the equivalent circuit depicted in Fig.1(a). Each junction is
characterized by a critical current $I_c $, resistance $R$, and
capacitance $C$; each loop has an inductance $L$. The two loops,
characterized by the Josephson phase
variables $\varphi _{1}$ and $\varphi _{2}$, are coupled through a capacitance $%
C_X $. The externally applied fluxes through loops 1 and 2 are represented
by $\Phi _{1x} $ and $\Phi _{2x} $. As discussed in [15], in dimensionless
form the governing equations of this system turn out to be:
\begin{eqnarray}
\ddot{\varphi}_{a}+\alpha \dot{\varphi}_{a}+\sin \varphi _{a}\cos \varphi
_{b}=-\beta _{L}^{-1}\left[ {\varphi _{a}+2\pi M_{a}}\right]
\label{eq1}\\
g^{-1}\ddot{\varphi}_{b}+\alpha \dot{\varphi}_{b}+\sin \varphi _{b}\cos \varphi
_{a}=-\beta _{L}^{-1}\left[ {\varphi _{b}+2\pi M_{b}}\right] \; ,
\label{eq2}
\end{eqnarray}
\noindent with overdots denoting derivatives in dimensionless time
$\tau =\omega _{J}t$. $\varphi _{a}=(\varphi _{1}+\varphi _{2})/2$ and $\varphi
_{b}=(\varphi _{1}-\varphi _{2})/2$ are the transformed phase variables
that are convenient for the analysis. The junction plasma frequency
is $\omega _{J}=\sqrt{2eI_{c}/\hbar C}$; $\alpha =1/\omega _{J}CR$
and $\beta _{L}=2\pi
LI_{c}/\Phi _{0}$, $\Phi _{0}=h/2e$ being the flux quantum. $%
g^{-1}=1+2\gamma _{x}$ with $\gamma _{x}=C_{x}/C$ representing the mutual
coupling coefficient. $M_{ix}=\Phi _{ix}/\Phi _{0}$ is the total normalized
applied flux through loop $i$; $M_{a}=\tfrac{1}{2}(M_{1x}+M_{2x})$ and $%
M_{b}=\tfrac{1}{2}(M_{1x}-M_{2x})$ are corresponding transformed normalized magnetic
fields.
As in [15], parameter values were set at $\alpha =5\times 10^{-5}$, $\beta
_{L}=2.841$, $g=0.9954$, $I_{c}=1.1\mu A$, $C=1.3pF$, and $\omega
_{J}^{-1}=0.02ns$. Both loops are biased with a dc flux $M_{ix}=0.6941$ (resonance $%
\approx 5.1$GHz) and with superimposed microwave (MW) pulses as shown in Fig.3(A) of
Ref.~[14].

Our previous work [15] outlined the dynamical modes exhibited by this system. One
of these modes is consistent with the experimental observations and can be
obtained in the linear limit of oscillation amplitude. Following Ref.~[15],
we therefore take $M_{b}=0$, $\alpha \approx 0$, and for small amplitude
oscillations of $\varphi _{1}$ and $\varphi _{2}$, we write $\varphi _{a}=\varphi
_{0}+\psi _{a}$, where $\varphi _{0}$ is a constant and $|\psi _{a}|\ll 1$.
Similarly we assume $|\varphi _{b}|\ll 1$. Inserting this ansatz with $\alpha =0
$ (for simplicity) gives the linear equations
\begin{eqnarray}
\beta _L \sin \varphi _0 + \varphi _0 + 2\pi M_a & = & 0 \label{eq3}\\
\ddot{\psi}_{a}+(\cos \varphi _{0}+\beta _{L}^{-1})\psi _{a} & = & 0
\label{eq4}\\
g^{-1}\ddot{\varphi}_{b}+(\cos \varphi _{0}+\beta _{L}^{-1})\varphi _{b} & = & 0 \; .
\label{eq5}
\end{eqnarray}
The first of these equations provides the average phase $\varphi _0 $
for each of the two loops given the external DC magnetic field. The
next two equations have solutions $\psi _a = A\sin (\omega _a t + \theta _a )$
and $\varphi _b = B\sin (\omega _b t + \theta _b )$,
where the four parameters, $A,B,$ $\theta _{a}$ and
$\theta _{b}$ are given by initial conditions. The two resonance
frequencies are $\omega _{a}^{2}=\cos \varphi _{0}+\beta _{L}^{-1}$ and
$\omega _{b}=\sqrt{g}\omega _{a}$. The Josephson phases $\varphi_i$
of the two loops therefore evolve according to
\begin{eqnarray}
\label{eq8}
&& \varphi_i-\varphi_0 \; =  \; \psi _a \pm \varphi _b \\
&& = \; \sqrt {A^2 + B^2 \pm 2AB\cos 2(\omega _d t + \theta _d )} \sin
(\omega _s t + \theta _i ) \; , \nonumber 
\end{eqnarray}
where $i=1,2$, and "$+$" and "$-$" apply to $i=1$ and $i=2$, respectively.
The frequencies and phases are
$\omega _s = \tfrac{1}{2}(\omega _a + \omega _b )$, $\omega
_d = \tfrac{1}{2}(\omega _a - \omega _b )$, and $\theta _d = \tfrac{1}{2}%
(\theta _a - \theta _b )$. This expression directly gives the
experimentally [14] and theoretically [15] observed slow envelope modulation of the
two coupled oscillators, and calculating the modulation frequency
from the experimentally provided system parameters yields $%
\Omega _I =2\omega_d\approx 0.001455$, which is in very good agreement with
observations [14,15]. The two phases $\theta _1 $ and $\theta _2 $
are also modulated with $\omega_d$ such that
\begin{eqnarray} 
\sin \theta _i & = & A\cos (\omega _d t + \theta _a ) \pm B\cos (\omega
_d t
- \theta _b )\label{eq10} \\
\cos \theta _i & = & - A\sin (\omega _d t + \theta _a ) \pm B\sin
(\omega _d t - \theta _b ) \; . \label{eq11}
\end{eqnarray}

We now discuss how these simple linear oscillations are related to the
Bloch-vectors that have been used in the literature to illustrate the system
behavior. In analogy with the quantum mechanical picture, we define a
Bloch-sphere centered in a Cartesian coordinate system with a horizontal $xy$%
-plane and a vertical z-axis. The corresponding Bloch vector of an
oscillator is given by the phase and amplitude of oscillation such that the
phase of oscillation provides the direction in the xy-plane and the
amplitude (or, equivalently, the energy) provides the $z$-coordinate. The
definition of the vertical axis is given by the switching probabilities when
the probe pulse is applied: $z = - 1$ (vertical down) is a state with
near-zero switching probability; $z = 1$ (vertical up) is a state with
near-unity switching probability; and $z = 0$ (horizontal) are states with intermediate
switching probability. In the present system of two coupled Josephson
loops, the $z=0$ states (ideally 50{\%} switching) are generated by initially
applying a $\pi$-MW pulse of half a Rabi-period duration (notice that
Rabi-type oscillations are already characterized in the classical system
[13]) to one of the loops, then allowing the two coupled oscillators to
exchange energy during a time of free evolution until their energy contents,
and therefore switching probabilities, coincide (see Refs.~[14,15]).
Notice that the experiments can only measure the vertical axis $z$ by the
switching probability. Thus, in order to experimentally gain any phase
information from an oscillator (i.e., the \textit{xy}-coordinate of the
Bloch vector in the \textit{xy}-plane), one must "rotate" the Bloch vector
$\tfrac{\pi}{2}$ around horizontal axes of the Bloch-sphere before conducting the
measurement [14]. This is done by applying additional $\tfrac{\pi}{2}$-MW pulses of a duration
that is about half of the initial $\pi$-MW pulse. The
phase-difference between such rotation pulse and the perturbed oscillator
determines if the oscillator energy is amplified ($z$ increases) or
attenuated ($z$ decreases). Applying pulses of different phases can
therefore illuminate the phase (the $xy$ coordinates) of a Bloch-vector by
subsequent switching ($z$) measurements. A pair of two such orthogonal $\tfrac{\pi}{2}$
pulses are denoted $X$ and $Y$. This is the classical analog to state
tomography.

We conduct state tomography by simultaneous switching measurements of the
two loops after having applied rotation pulses $(I,X,Y)$. We acquire the
probabilities in the form, where, e.g., $P_{X_1 Y_0 } $ is the probability
for simultaneously measuring a switch in loop 1 and a no-switch in loop 2
after having applied an $X$-rotation to loop 1 and a $Y$-rotation to loop 2.
The notation for not applying an $X$ or $Y$ rotation is $I$. Following the
quantum mechanical treatment, these measured probabilities can be fed into
expressions for the Hermitian density matrix $\{\rho _{jk} \}_{4\times4} $
in, e.g., the following way: $\rho _{11} = P_{I_0 I_0 }$,  $\rho _{22} = P_{I_0 I_1 }$,
$\rho _{33} = P_{I_1 I_0 }$, $\rho _{44} = P_{I_1 I_1 }$, and
\begin{eqnarray} 
2\rho _{12} & = & (P_{I_0 X_1 } - P_{I_0 X_0 } ) + 
i(P_{I_0 Y_0 } - P_{I_0 Y_1 } )\label{eq20}\\
2\rho _{13} & = & (P_{X_1 I_0 } - P_{X_0 I_0 } ) + 
i(P_{Y_0 I_0 } - P_{Y_1 I_0 } )\label{eq20}\\
2\rho _{24} & = & (P_{X_1 I_1 } - P_{X_0 I_1 } ) + 
i(P_{Y_0 I_1 } - P_{Y_1 I_1 } )\label{eq20}\\
2\rho _{34} & = & (P_{I_1 X_1 } - P_{I_1 X_0 } ) +
i(P_{I_1 Y_0 } - P_{I_1 Y_1 } )\label{eq21}\\
2\rho _{23} & = & (1 - P_{X_0 X_1 } - P_{X_1 X_0 } - P_{Y_0
Y_1 } - P_{Y_1 Y_0 } ) \label{eq23}\\
& + & i(P_{Y_0 X_1 } + P_{Y_1 X_0 } - P_{X_0 Y_1 } - P_{X_1
Y_0 } )\nonumber \\
2\rho _{14} & = & (P_{Y_0 Y_1 } + P_{Y_1 Y_0 } - P_{X_0 X_1 }
- P_{X_1 X_0 } ) \label {eq14}\\
& + & i(P_{X_0 Y_1 } + P_{X_1 Y_0 } + P_{Y_0 X_1 } + P_{Y_1
X_0 } - 1) \; . \nonumber
\end{eqnarray}
Our only deviation from the experimental work [14] is that we use the direct
expressions above instead of a least-square fit.


Simulations were conducted as follows.
Each loop was initially at rest in its shallow potential well
generated by the DC magnetic field as illustrated in Fig.1(b). A $\pi$-MW
pulse was applied with a randomly chosen phase and a normalized amplitude of
$7.5\times10^{-5}$ for 500 time units to loop 2. This elevates loop 2 to a
relatively high energy state, while loop 1 is left near the bottom of its
well. The free evolution of the system is therefore approximately given by
Eq.~(6) with $\theta _{d}\approx 0$ and $0\leq B\lesssim A$ after
the initial $\pi$-MW field on loop 2 is terminated. The two coupled oscillators
now evolve until a rotation pulse is applied to one or both
of the oscillators, after which possible switching is measured by applying a
probe pulse to each loop. The rotation pulses were chosen in accordance with
the experiments with a duration of 200 time units and same amplitude as the
initial $\pi$-MW pulse. Their phases were randomly chosen, but such that $X$ and $Y$
pulses were $\tfrac{\pi}{2}$ out of phase. The probe pulse was calibrated so
that the initial state prior to MW pulses would yield vanishing switching
probability, the state of loop 2 immediately following the initial $\pi$-MW
pulse would result in near certain switching, and the states, where the
energies of the two loops become equal, would result in some intermediate
percentage of significance. Thus, we chose a
half-sine wave with duration 500 time units and normalized amplitude $%
3.6\times10^{-3}$. The energies of the two loops became equal at a time of
approximately 900 units after the termination of the $\pi$-MW pulse, and
the probe pulse was centered near this value. The $(I),X,Y$ rotation pulses
were applied immediately prior to the application of the probe pulse. We
repeated such simulations 2000 times with different values of the randomly
chosen phases in order to obtain a set of switching probabilities for any
given combination of rotation pulses. We then generated
a density matrix from the simultaneous switching probabilities, as outlined above.
A typical example of this is presented in Figure
2, where the real and imaginary parts of the matrix are shown separately. We
note the striking resemblance to the density matrix that was previously
published (Fig.~3B of Ref.~[14]) based on the experimental data. 
The matrix shows relatively small real
magnitudes outside the diagonal, and relatively small imaginary elements, except
for $\rho_{23}$, which contain the signature of "entanglement".
Thus, simulation data generated from the classical RCSJ model yields a
density matrix that matches in its essential features the density matrix
produced from experimental observations of the coupled qubit system. In
order to emphasize the similarity between our result here and that of Ref.
[14], we can calculate the \textquotedblleft fidelity\textquotedblright\ $%
F_{\exp }=\sqrt{\nu ^{\ast T}\rho _{data}\nu }$, where $\rho _{data}$ is the
density matrix developed based on the acquired classical data and $\nu =(0,\tfrac{1}{
\sqrt{2}},-\tfrac{i}{\sqrt{2}},0)^{T}$ is the vector corresponding to the ideally
expected entangled state [5]. With the developed density matrix shown in
Figure 2 we get $F_{\exp}=0.77$. This value is very close to the one obtained
in Ref. [14], and we will now discuss this result.

\begin{figure}
[pt]
\begin{center}
\includegraphics[
trim=1.0in 6.0in 0.0in 0.0in,
width=3.50in
]%
{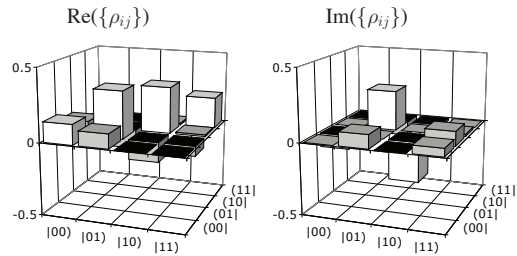}%
\caption{State tomography of ``entangled'' qubits. Real and
imaginary components, respectively, of the density matrix. We
deliberately choose the notation $\vert ij)$ rather than $\vert ij
 \rangle$ to distinguish our results as fundamentally classical
rather than quantum mechanical. This density matrix should be compared to Fig.~3B of Ref.~[14].}%
\label{Fig.2}%
\end{center}
\end{figure}


The probe pulse was calibrated to give approximately the same
optimal switching probability for the two loops when no rotation
pulses were applied. The implication of this is that the
corresponding Bloch vectors would be positioned in the xy-plane of
the Bloch sphere when these simultaneous measurements are conducted.
The probability $P_{1}$ for switching a loop can therefore be
related to the z-coordinate of the unit-Bloch-vector in the simple
way $P_{1}=\tfrac{1}{2} (1+z)$. Correspondingly, if an $X$ or a $Y$
rotation is applied prior to the measurement, one would have the
probabilities $P_{X_1}=\tfrac{1}{2} (1+\sin \theta )$\textit{\ }and
$P_{Y_1}=\tfrac{1}{2} (1-\cos \theta )$, where $\theta $ is the angle
(relative to the $x$-axis) of the Bloch-vector in the xy-plane prior to the application of a
$\tfrac{\pi}{2}$ rotation pulse. Notice that the probability of switching in
the classical system is given only by the sampling of the outcomes
of random MW-phases (at the extreme low temperature limit), whereas
the quantum mechanical picture sees the switching
probability as only due to tunneling. In averaging over all random phases $%
\theta $, one can get a good sense of what to expect from the
density matrix elements. For example the particular element $\rho
_{23}$ corresponding to a random-phase averaged ensemble of two
Bloch-vectors, which are $\Gamma=\theta_1-\theta_2$ out of phase in the xy-plane, we
have $\rho _{23}=\tfrac{1}{4} \exp (i\Gamma )$. This explains the
information that the density matrix provides. The important
element $\rho _{23}$ is a measure of the angle $\Gamma$ between the two
simultaneously detected
Bloch-vectors in the xy-plane -- or, equivalently, the phase difference
between the Josephson phase oscillations of the two loops. If we produce an element $\rho
_{23}\approx\pm\tfrac{i}{4}$ (which is what is obtained), then this arises from
two vectors that are nearly orthogonal in the
\textit{xy}-plane -- or, equivalently, that the oscillations of the
two loops are $\tfrac{\pi}{2}$ out of phase. However, the formalism does not
explain {\it why} this is, and the result may have nothing to do with
quantum mechanics, as we have seen here.

The fact that the classical RCSJ model produces this particular result is
not a coincidence. We concluded above that the $\pi$-MW initiation of the system provides for
the parameters $0\leq B\lesssim A$ and $\theta _{d}\approx 0$. If we
observe the system at the time $t_{m}$ when the two energies coincide
($\omega _{d}t_{m}=\tfrac{\pi}{4}$), we can see from equations (7) and (8) that $\Gamma
=\theta _{1}-\theta _{2}=\tan ^{-1}\tfrac{A-B}{A+B}-\tan ^{-1}\tfrac{A+B}{A-B}$,
which is $\Gamma =-\tfrac{\pi}{2}$ for $A=B$, and $\Gamma \gtrsim -\tfrac{\pi}{2}$
for $A\gtrsim B\geq 0$. Thus, the two coupled oscillators are naturally
orthogonal in phase when their energy contents are similar, and the resulting
density matrix, obtained from the switching measurements after rotations, therefore
reflects this fact.

We have demonstrated that the classical RCSJ model of two coupled qubits exhibits
the same signatures as observed experimentally when conducting state tomography
and examining the resulting density matrix. These signatures have previously been used
to argue for an unambiguous demonstration of quantum entanglement. However, given that
our data are definitely classical in origin, we submit that such definite conclusions
cannot be made. Instead, the density matrix, and the element $\rho_{23}$ in particular,
provides information only about the phase difference between the Josephson oscillations
of the two superconducting loops without providing the underlying reason for this phase
difference. Assuming that the system is governed by quantum mechanics this may be
interpreted as a signature of quantum entanglement. However, one does not need to
assume this, and our classical analysis of the system has revealed that exactly the same
signature in the density matrix appears as a result of the weak coupling.
This leads to an {\it ambiguous} interpretation of the observed phenomena.
We therefore conclude that unambiguous
demonstrations of quantum behavior in this class of systems must be analyzed
not only in light of the quantum mechanical model and its expectations, but
also in light of what the corresponding classical model can predict and
explain. 

\begin{acknowledgments}
We are grateful for very useful discussions with Massimo Bianchi, Maria
Gabriella Castellano, Mogens R.~Samuelsen, and Stuart A.~Trugman. One of us
(J.A.B.) received financial support from the Natural Sciences
and Engineering Research Council of Canada.
\end{acknowledgments}






\begin{thebibliography}{99}                                                                                        %

\bibitem {Bardeen}J. Bardeen, Phys. Rev. Lett. 42, 1498 (1979); R. E. Thorne, J. R. Tucker,
and J. Bardeen, Phys. Rev. Lett. \textbf{58}, 828 (1987).

\bibitem {Sneddon}L. Sneddon, M. C. Cross, and D. S. Fisher, Phys. Rev. Lett. \textbf{49}, 292
(1982); R. A. Klemm and J. R. Schrieffer, Phys. Rev. Lett. \textbf{51}, 47
(1983).

\bibitem {Pancharatnam}S. Pancharatnam, Proceedings of Indian Academic of Science, \textbf{44}, A,
247 (1956).M. V. Berry, Proceedings of the Royal Society of London, A,
\textbf{392}, 45 (1984).

\bibitem {Chao}R. Y. Chiao, Phys. Rev. Lett. \textbf{57}, 933 (1986); A. Tomita and R. Y.
Chiao, Phys. Rev. Lett. \textbf{57}, 937 (1986)

\bibitem {Haldane}F. D. Haldane , Phys. Rev. Lett \textbf{59}, 1788 (1987)

\bibitem {Segertt}J. Segert, Phys. Rev. A \textbf{36}, 10 (1987)

\bibitem {VanDuzer}T. Van Duzer and C.W. Turner, \textit{Principles of Superconductive Devices
and Circuits} (Elsevier, New York,1981), Chapter 5.

\bibitem {Anderson}P. W. Anderson, in \textit{Lectures on the Many Body Problem}, Ed. by E. R.
Caianiello (Academic Press NY 1964), Vol. 2, p. 132.

\bibitem {Caldeira}A. O. Caldeira and A. J. Leggett, Phys. Rev. Lett. \textbf{46}, 211 (1981);
A. J. Legget and A. Garg, Phys. Rev. Lett. \textbf{54}, 857 (1985).

\bibitem {affleck}I. Affleck, Phys. Rev. Lett. \textbf{46}, 388 (1981).

\bibitem {Martinis}J.M. Martinis, M.H. Devoret, and J. Clarke, Phys. Rev. Lett. \textbf{55},
1543 (1985).

\bibitem {list1}See, e.g., J.M. Martinis, S. Nam, J. Aumentado, and C. Urbina,
Phys. Rev. Lett. {\bf 89}, 117901 (2002); R.W. Simmonds, K.M. Lang, D.A. Hite, S.
Nam, D.P. Pappas, and J.M. Martinis, Phys. Rev. Lett. {\bf 93}, 077003 (2004); J.
Claudon, F. Balestro, F.W.J. Hekking, and O. Buisson, Phys. Rev. Lett. {\bf 93}
(2004); D. Vion, A. Aassime, A. Cottet, P. Joyez, H. Pothier, C. Urbina, D.
Esteve, and M.H. Devoret, Forschr. Phys. {\bf 51}, 462 (2003).

\bibitem {list2}N. Gr{\o }nbech-Jensen {\it et al.}, Phys. Rev. Lett. {\bf 93}, 107002 (2004);
N. Gr{\o }nbech-Jensen and M. Cirillo, Phys. Rev. Lett., \textbf{95} 067001 (2005); J.
E. Marchese, M. Cirillo, and N. Gr{\o }nbech-Jensen, Phys. Rev. B \textbf{73}%
, 174507 (2006); N. Gr{\o }nbech-Jensen {\it et al.}, ``Anomalous thermal
escape in Josephson systems perturbed by microwaves'', in
\textit{Quantum Computing:
Solid State Systems} eds. B.~Ruggiero, P.~Delsing, C.~Granata, Y.~Paskin,
and P.~Silvestrini (Springer, New York, 2006), pp. 111-119; J. E. Marchese,
M. Cirillo, and N. Gr{\o }nbech-Jensen, Eur. Phys. J. \textbf{147}, 333-342
(2007).

\bibitem {Steffen}M. Steffen, M. Ansmann, R.C. Bialczak, N. Katz, E Lucero, R. McDermott, M.
Neeley, E.M. Weig, A.N. Cleland, J.M. Martinis, Science \textbf{313},
1423 (2006).

\bibitem {Blackburn}J.A. Blackburn, J.E. Marchese, M. Cirillo, and N. Gr{\o }nbech-Jensen, Phys.
Rev. \textbf{B 79}, 054516 (2009).

\end{thebibliography}
\end{document}